\documentclass[preprint]{aastex}

\usepackage{booktabs}
\usepackage{multirow}
\usepackage{graphicx}
\usepackage{hyperref}

\begin{document}

\title{\emph{Herschel} far-infrared photometric monitoring of protostars \\in the Orion Nebula Cluster}

\shorttitle{Far-infrared photometric monitoring of protostars}


\author{N. Billot}
\affil{NASA Herschel Science Center, California Institute of Technology, \\770 S. Wilson Ave, Pasadena, CA 91125, USA, \\and \\
Instituto de Radio Astronom\'ia Milim\'etrica, \\
Avenida Divina Pastora, 7, Local 20, 18012 Granada, Spain}
\email{billot@iram.es}

\author{M. Morales-Calder\'{o}n, J.~R. Stauffer}
\affil{Spitzer Science Center, Caltech, Pasadena, CA 91125, USA}

\author{T. Megeath}
\affil{University of Toledo, Department of Physics and Astronomy, Toledo, OH 43606, USA}

\author{B. Whitney}
\affil{Space Science Institute, Boulder, CO 80301, USA}

\begin{abstract}

We have obtained time series observations of the Orion Nebula Cluster at 70$\,\mu$m and 160$\,\mu$m from the \emph{Herschel}/PACS Photometer. This represents the first wide-field far-infrared photometric monitoring of a young star forming region. The acquired 35\arcmin$\times$35\arcmin~maps show complex extended structures, with unprecedented details, that trace the interaction between the molecular gas and the young hot stars. We detect 43~protostars, most of which are situated along the integral-shaped filament extending from the Orion nebula, through OMC$\,$2 and to OMC$\,$3. We present high-reliability light curves for some of these objects using the first six epochs of our observing program spread over 6~weeks. We find amplitude variations in excess of 20\% for a fraction of the detected protostars over periods as short as a few weeks. This is inconsistent with the dynamical time-scales of cool far-IR emitting material that orbits at hundreds of AU from the protostar, and it suggests that the mechanism(s) responsible for the observed variability originates from the inner region of the protostars, likely driven by variable mass accretion.

\end{abstract}


\section{Introduction}
\label{sec:intro}

Photometric variability was one of the original, defining characteristics of the class of objects that were later determined to be stars in the process of formation, or Young Stellar Objects (YSOs) \citep{joy}. While optical and near-IR variability studies have demonstrated the relation between accretion shocks and hotspots on the surface of rotating YSOs \citep[e.g.,][]{vrba, carpenter}, observations at longer wavelengths offer another perspective to study cooler circumstellar disks and envelopes surrounding nascent stars. In particular, recent mid-IR photometric and spectroscopic time series of YSOs with the \emph{Spitzer} Space Telescope \citep{werner} have begun to shed light on inner disk structures, identifying both inner disk warps and `clouds' in the disk \citep[e.g.,][]{morales09, muzerolle, espaillat}. Elaborate models offer satisfying explanations to the observed time series \citep{flaherty, turner, flaherty11, ke}.

Similar photometric variability studies are a lot sparser in the literature at longer wavelengths ($\lambda > 50\,\mu$m), mostly due to the difficult access to this wavelength domain. \citet{harvey} report on a 200\% flux variation of SSV$\,$13 in NGC$\,$1333 over a two-year period using the Kuiper Airborne Observatory. \citet{juhasz} and \citet{sitko} analyze the 1-300$\,\mu$m variability of the star SV$\,$Cep over two years, and they invoke a growing warp at the inner edge of the disk, which leads to the shadowing of the outer disk, to reproduce the observations across the entire wavelength range. Based on reprocessed ISOPHOT observations \citep{lemke}, \citet{kospal} present the far-IR fading of the pre-main sequence star OO$\,$Serpentis after it went into outburst, and they show that the $60\,\mu$m flux changed by a factor of 8 over a period of 11~years. Most sources targeted by these monitoring programs are disk-bearing YSOs.

In this article, we present the first sensitive wide-field far-IR photometric monitoring of protostars which significantly extends the statistics on YSO variability monitoring in this wavelength regime and evolutionary stage. We recently completed the first set of time series observations of the Orion Nebula Cluster (ONC) with the PACS instrument \citep{poglitsch} onboard the \emph{Herschel} Space Observatory \citep{pilbratt}. During the first two-month visibility window, we obtained six observations of a single field centered on the integral-shaped filament located north of the ONC, which contains hundreds of YSOs. 
In addition, this field has been monitored in the optical \citep{herbst}, and in the mid-IR with \emph{Spitzer} as part of the YSOVAR program \citep{morales11}, thus making it the largest YSO variability database to date. 

We first describe the observations and data reduction in section~\ref{sec:method}, then we give a detailed account of the photometric measurements and present the light-curves extracted from the time series observations. Finally we discuss the possible origin of the observed variability in section~\ref{sec:result}.


\section{Method}
\label{sec:method}

\subsection{Observations}
\label{sec:obs}

We use the \emph{Herschel/}PACS Photometer simultaneously in the blue and red channels, operating at 70 and 160$\,\mu$m respectively, to measure photometric time series of YSOs in the ONC. This \emph{Herschel}~OT1 observing program is designed to make the best use of the observatory while accommodating the constraints dictated by the need for precise photometric monitoring.

The building block of our program is a 30-minute single-direction scan map observation covering a 35\arcmin$\times$35\arcmin~field. This scheme of scanning a relatively large area that contains several YSOs was preferred over repeated observations of individual sources because of the prohibitive telescope overheads that this would entail. The targeted field was chosen to contain as many YSOVAR-monitored protostars as possible, with over 100~known in total \citep{kryukova}. This building block is then repeatedly scheduled with absolute timing constraints to cover several 2-months visibility windows spread over a period of nearly 2~years. Moreover, within each PACS-Photometer observing campaign, which lasts 2.5-days, our observations are scheduled at least 12~hours after the end of the cryo-cooler recycling and switch-on of the detectors. This is to ensure the highest thermal and electrical stability possible required for photometric variability monitoring. 

In addition, considering that map-making algorithms rely on scan-direction redundancy to filter out instrumental drifts from the extended emission present in the field \citep[e.g.,][]{cantalupo, roussel}, we require that the scan direction of consecutive observations are rotated by 90\degr~with respect to the previous one. The scan-direction redundancy is further increased by the secular rotation of the spacecraft roll angle along its orbit ($\sim$10\degr~between consecutive observations). This observing strategy thus allows a very good rejection of instrumental drifts when combining single-epoch observations, and consequently the reconstruction of artifact-free extended emission structures\footnote{A Spitzer - Herschel composite image based on this dataset has also been released on the NASA web site at \href{http://www.nasa.gov/mission_pages/herschel/multimedia/pia13959.html}{http://www.nasa.gov/mission\_pages/herschel/multimedia/ pia13959.html}} (see Figure~\ref{fig:maps}).

\subsection{Data Processing}
\label{sec:processing}

We retrieve the raw data from the \emph{Herschel} Science Archive and process individual observations to obtain calibrated data cubes (\emph{Level$\,$1} frames) using the PACS standard pipeline with the \emph{Herschel} Interactive Processing Environment \citep[HIPE version~7.0~1931,][]{ott}. We correct the bolometer signals for non-linearities based on pre-launch calibration measurements \citep{billot06}. We also modify the frame's astrometry to correct for telescope pointing errors (up to 3\arcsec~in the sixth visit). 

We create maps from individual single-direction scan observations using the standard approach of highpass filtering the bolometer timelines and projecting the data cube on regular grids with pixel sizes of 1\arcsec~and 2\arcsec~in the blue and red PACS channels, respectively. We use a filter width twice as long as a scan leg (2000 readouts), and we mask bright regions prior to applying the filter to minimize shadowing artifacts along the scan direction. This filtering process alters to some extent the largest spatial scales in the map but preserves the signal at the scale of point sources. The source photometry is measured from these single-epoch maps (see section~\ref{sec:photometry}). 

\begin{figure*}[t]
  \centering
    \begin{tabular}{c}
      \includegraphics[width=0.60\textwidth]{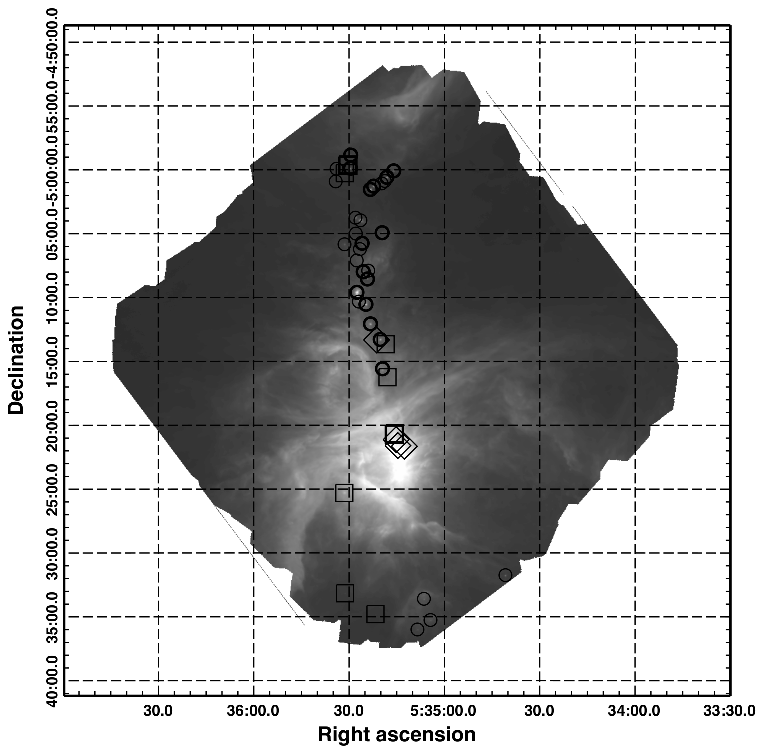}\\
      \includegraphics[width=0.60\textwidth]{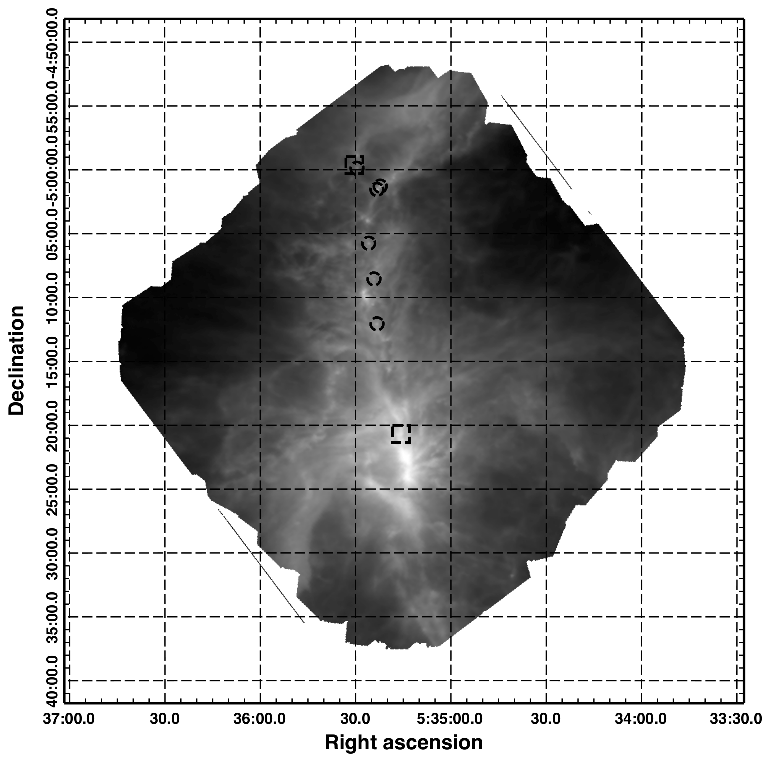}
    \end{tabular}
  \caption{Maps of the Orion$\,$A molecular ridge in the PACS 70$\,\mu$m (top) and 160$\,\mu$m (bottom) bands displayed on a logarithmic scale. The first six visits to the targeted field were combined to create these artifact-free maps. Symbols on the top image indicate the location of detected protostars from table~\ref{tab:fluxesAverage}: squares are Class$\,$II, circles Class$\,$I, and diamonds unclassified by the YSOVAR program. Thick symbols designate protostars with reliable light curves from table~\ref{tab:fluxes_LC}. Symbols on the bottom image show variable protostars following the definition given in section~\ref{sec:result}.}
  \label{fig:maps}
\end{figure*}

In parallel, we export individual Level$\,$1 frames out of HIPE to be combined by the IDL-based map-making algorithm Scanamorphos \citep{roussel}. Figure~\ref{fig:maps} presents the maps resulting from the combination of the six epochs observed during the first visibility window between February 26, 2011 and April 10, 2011. Most of the YSOs in this field are situated along the Orion$\,$A molecular clouds OMC-1, -2 and -3, north of the ONC. A small area of 10\arcsec~radius is saturated in both maps at the location of the Trapezium cluster where the signal reaches $\sim4\times10^6\,$MJy$\cdot$sr$^{-1}$ and $\sim2\times10^5\,$MJy.sr$^{-1}$ in the blue and red channels, respectively. The Orion bar is visible to the south east of the Trapezium cluster. A detailed discussion of the extended emission structure in this field is beyond the scope of this letter and will be treated in a subsequent article.

\subsection{Source photometry}
\label{sec:photometry}

The extended emission in the ONC is highly structured (cf Figure~\ref{fig:maps}), and it exhibits a great dynamic range, in excess of 1:10000. This makes the fine tuning of source-finding algorithms very difficult to converge to a satisfying solution. We therefore opt for a more pragmatic visual inspection of the combined 70$\,\mu$m map to identify point-like sources. We find 43~objects, with fluxes ranging from 0.4 to 450$\,$Jy. About 80\% of these sources have 160$\,\mu$m counterparts. One third of those appear to be point-like objects at 160$\,\mu$m, with fluxes ranging from 13 to 280$\,$Jy, while the other two thirds are spatially resolved and often indistinguishable from the underlying extended emission structure. Most of the sources detected in the PACS maps have been previously identified as Class$\,$I/0 protostars by \citet{megeath12}, as presented in Table~\ref{tab:fluxesAverage}.

For each single-epoch map, we measure the photometry of point-like sources within a small aperture centered on the PSF core (4\arcsec~and 8\arcsec~radius in the blue and red bands, respectively). We measure background levels, and associated background standard deviations, within an annulus of inner/outer radii of 6\arcsec/8\arcsec~and 10\arcsec/16\arcsec~in the blue and red bands, respectively. Although the signal in the PSF wings affects to some extent the determination of the background level and standard deviation, it is best to work with small aperture/annulus radii to minimize the effect of the structured background emission. We correct the measured fluxes with aperture correction factors derived from the PACS~PSF\footnote{The PSFs are measured on the Asteroid Vesta, and are publicly available through the \emph{Herschel} Science Center web pages at http://herschel.esac.esa.int/.}. The PACS Photometer bolometer arrays are extremely stable detectors with gain variations around 1\% over a year timescale \citep{billot10}. Flux uncertainties are thus mainly dominated by errors in background level estimates, which are likely underestimated by the standard deviation measured in the thin annulus. 

We estimate empirical flux uncertainties as the typical flux variations measured on a given source in various combinations of aperture/annulus radii. At 70$\,\mu$m, we find that flux uncertainties are generally below 5\%, but they can reach 20\% for sources located in regions of highly structured backgrounds. At 160$\,\mu$m, the absolute flux uncertainties are generally higher ($\sim$20-50\%), mainly due to the structured background and to the fact that some sources are extended at this wavelength in which case aperture corrections are no longer appropriate and may introduce systematic errors. Table~\ref{tab:fluxesAverage} gives the source fluxes averaged over the visibility window period. The errors quoted in the table reflect the standard deviation measured in the background annulus and likely underestimate the flux uncertainties as described above.


Similarly, we estimate the relative flux uncertainties, i.e. the flux variations measured on a given source at different epochs, by exploring various combinations of aperture/annulus radii. Presently, if any point of a light curve shows a flux variation of more than 5\% in the various combinations of aperture/annulus radii explored, the light curve is deemed inaccurate. This rather stringent criterion ensures the highest reliability of the light curves. In the sample of 43~sources, we find that 17~have a reliable light curve in the blue band, and only 6~in the red band. Table~\ref{tab:fluxes_LC} gives the time series fluxes of sources that pass this selection criterion, and figure~\ref{fig:lightCurves} presents a sample of these high-reliability light curves.

\begin{deluxetable}{lccccc}\centering
\tablecolumns{9}
\tabletypesize{\scriptsize}
\tablewidth{0pt}
\tablecaption{Source flux averaged over the visibility window period. \label{tab:fluxesAverage}}
\tablehead{\colhead{Source ID} & \colhead{R.A.} & \colhead{Dec.} & \multicolumn{2}{c}{Flux (Jy)} & \colhead{YSO} \\
\colhead{} & \colhead{(deg.)} & \colhead{(deg.)} & \colhead{70\,$\mu$m} & \colhead{160\,$\mu$m\tablenotemark{a}} & \colhead{Class\tablenotemark{b}}}
\startdata
HOY J053529.44-045851.6 & 83.872 & -4.981 & 6.33$\pm$0.07 & 27.4$\pm$0.69 & I/0 \\   
HOY J053533.95-045955.1 & 83.891 & -4.998 & 0.58$\pm$0.01 & \emph{:::}  & I/0 \\   
HOY J053534.22-050054.2 & 83.892 & -5.015 & 2.13$\pm$0.02 & \emph{:::}  & I/0 \\  
HOY J053530.33-045938.6 & 83.876 & -4.994 & 0.93$\pm$0.01 & ... & II \\  
HOY J053531.36-050015.9 & 83.880 & -5.004 & 0.48$\pm$0.01 & ...  & II  \\  
HOY J053529.50-045952.7 & 83.872 & -4.997 & 12.9$\pm$0.10 & 21.6$\pm$0.69  & I/0 \\  
HOY J053515.98-050004.7 & 83.816 & -5.001 & 7.29$\pm$0.08 & \emph{:::}  & I/0 \\   
HOY J053518.12-050035.6 & 83.825 & -5.009 & 27.4$\pm$0.22 & 38.7$\pm$1.60  & I/0 \\   
HOY J053518.73-050053.1 & 83.828 & -5.014 & 2.82$\pm$0.03 & \emph{:::}  & I/0 \\  
HOY J053519.74-050104.7 & 83.832 & -5.017 & 2.28$\pm$0.05 & \emph{:::}  & I/0 \\  
HOY J053522.27-050116.8 & 83.842 & -5.021 & 33.5$\pm$0.28 & 70.4$\pm$1.63  & I/0 \\  
HOY J053523.26-050132.6 & 83.846 & -5.025 & 52.5$\pm$0.38 & 158$\pm$2.58 & I/0 \\   
HOY J053528.00-050343.1 & 83.866 & -5.061 & 18.0$\pm$0.21 & 27.2$\pm$1.39  & I/0 \\  
HOY J053526.43-050356.5 & 83.860 & -5.065 & 72.1$\pm$0.75 & 73.1$\pm$1.61  & I/0 \\  
HOY J053519.61-050453.7 & 83.831 & -5.081 & 3.60$\pm$0.06 & ...   & I/0 \\   
HOY J053531.47-050549.1 & 83.881 & -5.096 & 5.63$\pm$0.06 & \emph{:::}  & I/0 \\  
HOY J053527.93-050459.0 & 83.866 & -5.083 & 2.40$\pm$0.02 & \emph{:::}  & I/0 \\  
HOY J053525.89-050544.2 & 83.857 & -5.095 & 12.3$\pm$0.11 & 41.6$\pm$1.02  & I/0 \\   
HOY J053526.58-050610.7 & 83.860 & -5.102 & 6.80$\pm$0.05 & \emph{:::}   & I/0\\  
HOY J053525.51-050759.1 & 83.856 & -5.133 & 10.5$\pm$0.09 & \emph{:::}  & I/0 \\   
HOY J053527.56-050704.6 & 83.864 & -5.117 & 1.39$\pm$0.11 & \emph{:::}  & I/0 \\ 
HOY J053523.95-050753.7 & 83.849 & -5.131 & 2.88$\pm$0.07 & \emph{:::}  & I/0 \\ 
HOY J053524.23-050831.9 & 83.850 & -5.142 & 7.23$\pm$0.10 & 18.2$\pm$1.33  & I/0 \\   
HOY J053524.71-051030.6 & 83.852 & -5.175 & 95.0$\pm$1.21 & 36.6$\pm$4.19  & I/0 \\   
HOY J053526.91-051017.5 & 83.862 & -5.171 & 17.5$\pm$0.79 & ...  & I/0 \\ 
HOY J053527.56-050934.5 & 83.864 & -5.159 & 450$\pm$6.51 & 279$\pm$9.42  & I/0 \\   
HOY J053523.29-051203.0 & 83.847 & -5.200 & 48.4$\pm$0.57 & 54.7$\pm$2.16  & I/0 \\   
HOY J053520.15-051317.4 & 83.833 & -5.221 & 46.8$\pm$0.38 & \emph{:::}  & I/0 \\   
HOY J053521.34-051319.5 & 83.838 & -5.222 & 6.26$\pm$0.09 & \emph{:::} & -- \\   
HOY J053518.50-051340.2 & 83.827 & -5.227 & 3.68$\pm$0.12 & ...  & II  \\    
HOY J053519.47-051534.7 & 83.831 & -5.259 & 27.1$\pm$0.59 & \emph{:::}  & I/0 \\   
HOY J053517.93-051614.8 & 83.824 & -5.270 & 3.51$\pm$0.13 & ...  & II  \\  
HOY J053531.42-052515.7 & 83.880 & -5.421 & 31.9$\pm$0.98 & \emph{:::}  & II  \\ 
HOY J053515.69-052040.6 & 83.815 & -5.344 & 189$\pm$6.60 & \emph{:::}  & II  \\   
HOY J053514.71-052135.9 & 83.811 & -5.359 & 130$\pm$8.48 & ... & -- \\   
HOY J053515.26-052109.6 & 83.813 & -5.352 & 117$\pm$6.73 & \emph{:::} & -- \\  
HOY J053512.71-052139.5 & 83.802 & -5.360 & 216$\pm$12.9 & ... & -- \\ 
HOY J053531.30-053308.9 & 83.880 & -5.552 & 0.92$\pm$0.02 & \emph{:::}  & II  \\  
HOY J053521.65-053447.1 & 83.840 & -5.579 & 0.38$\pm$0.01 & \emph{:::}  & II  \\ 
HOY J053506.50-053335.0 & 83.776 & -5.559 & 5.10$\pm$0.17 & ...  & I/0 \\  
HOY J053508.52-053558.4 & 83.785 & -5.599 & 2.62$\pm$0.05 & \emph{:::}  & I/0\\  
HOY J053504.19-053512.1 & 83.768 & -5.587 & 1.37$\pm$0.11 & \emph{:::}  & I/0 \\  
HOY J053440.92-053144.8 & 83.669 & -5.529 & 7.70$\pm$0.05 & 13.3$\pm$0.27  & I/0 \\  
\enddata
\tablenotetext{a}{The two symbols \emph{:::} and ... indicate that a source is extended or not detected, respectively.}
\tablenotetext{b}{Spectral energy distribution class following the scheme of \citet{greene}.}
\end{deluxetable}

\begin{deluxetable}{lccccccc}\centering
\tablecolumns{9}
\tabletypesize{\footnotesize}
\tablewidth{0pt}
\tablecaption{High-reliability flux time-series measured at 70$\,\mu$m and 160\,$\mu$m between February and April 2011. \label{tab:fluxes_LC}}
\tablehead{\colhead{Source ID} &\multicolumn{6}{c}{Flux (Jy)} & \colhead{$\Delta$\,Flux\tablenotemark{a}}\\ \cline{2-7}
\colhead{} & \colhead{Feb.~26} & \colhead{Mar.~6} & \colhead{Mar.~14} & \colhead{Mar.~21} & \colhead{Mar.~31} & \colhead{Apr.~10} & \colhead{(\%)}}
\startdata
\cline{1-8}
\multicolumn{8}{c}{PACS 70\,$\mu$m}\\ \cline{1-8}
HOY J053529.44-045851.6 & 6.36$\pm$0.07 & 6.21$\pm$0.07 & 6.20$\pm$0.05 & 6.58$\pm$0.06 & 6.35$\pm$0.05 & 6.29$\pm$0.05 & 6 \\
HOY J053530.33-045938.6 & 0.96$\pm$0.01 & 0.90$\pm$0.01 & 0.85$\pm$0.01 & 0.96$\pm$0.01 & 0.98$\pm$0.01 & 0.91$\pm$0.01 & \emph{15} \\
HOY J053529.50-045952.7 & 12.8$\pm$0.10 & 13.4$\pm$0.09 & 12.2$\pm$0.09 & 12.6$\pm$0.10 & 13.2$\pm$0.09 & 13.2$\pm$0.09 & \emph{10} \\
HOY J053515.98-050004.7 & 7.34$\pm$0.07 & 7.44$\pm$0.07 & 7.20$\pm$0.06 & 7.06$\pm$0.07 & 7.19$\pm$0.08 & 7.50$\pm$0.08 & 5 \\
HOY J053518.12-050035.6  & 27.3$\pm$0.17 & 27.5$\pm$0.22 & 27.5$\pm$0.19 & 27.2$\pm$0.19 & 28.2$\pm$0.21 & 26.4$\pm$0.18 & 7 \\
HOY J053522.27-050116.8 & 31.0$\pm$0.19 & 31.3$\pm$0.22 & 32.3$\pm$0.21 & 35.3$\pm$0.25 & 37.0$\pm$0.28 & 34.2$\pm$0.22 & \emph{19} \\
HOY J053523.26-050132.6 & 51.8$\pm$0.32 & 53.5$\pm$0.36 & 49.6$\pm$0.32 & 50.5$\pm$0.38 & 55.1$\pm$0.38 & 54.7$\pm$0.33 & \emph{12} \\
HOY J053519.61-050453.7 & 3.58$\pm$0.04 & 3.48$\pm$0.04 & 3.49$\pm$0.04 & 3.58$\pm$0.04 & 3.77$\pm$0.06 & 3.70$\pm$0.05 & 8 \\
HOY J053525.89-050544.2 & 11.2$\pm$0.09 & 12.4$\pm$0.11 & 12.5$\pm$0.10 & 11.7$\pm$0.11 & 12.8$\pm$0.10 & 13.0$\pm$0.10 & \emph{17} \\
HOY J053525.51-050759.1 & 10.5$\pm$0.07 & 10.6$\pm$0.08 & 10.5$\pm$0.07 & 10.0$\pm$0.09 & 10.5$\pm$0.07 & 10.7$\pm$0.09 & 7 \\
HOY J053524.23-050831.9 & 7.61$\pm$0.08 & 7.93$\pm$0.09 & 7.25$\pm$0.10 & 6.53$\pm$0.10 & 7.04$\pm$0.09 & 7.05$\pm$0.10 & \emph{21} \\
HOY J053524.71-051030.6 & 93.6$\pm$0.90 & 93.4$\pm$1.00 & 95.5$\pm$1.13 & 94.9$\pm$1.21 & 97.7$\pm$1.15 & 94.7$\pm$1.15 & 7 \\
HOY J053527.56-050934.5 & 460$\pm$6.08 & 459$\pm$5.98 & 457$\pm$5.99 & 460$\pm$5.48 & 431$\pm$6.51 & 432$\pm$5.86 & 8 \\
HOY J053523.29-051203.0 & 50.7$\pm$0.44 & 49.5$\pm$0.57 & 46.2$\pm$0.39 & 46.1$\pm$0.38 & 50.7$\pm$0.48 & 47.0$\pm$0.52 & \emph{11} \\
HOY J053520.15-051317.4 & 45.2$\pm$0.34 & 46.7$\pm$0.32 & 46.2$\pm$0.30 & 47.3$\pm$0.35 & 48.1$\pm$0.30 & 47.0$\pm$0.38 & 7 \\
HOY J053519.47-051534.7 & 27.0$\pm$0.57 & 27.0$\pm$0.59 & 26.4$\pm$0.55 & 28.2$\pm$0.59 & 27.1$\pm$0.56 & 27.0$\pm$0.52 & 7 \\
HOY J053515.69-052040.6 & 191$\pm$6.43 & 181$\pm$6.60 & 190$\pm$6.27 & 190$\pm$5.83 & 204$\pm$6.45 & 180$\pm$6.45 & \emph{17} \\
\cline{1-8}
\multicolumn{8}{c}{PACS 160\,$\mu$m}\\ \cline{1-8}
HOY J053529.50-045952.7 & 20.9$\pm$0.68 & 22.6$\pm$0.66 & 21.7$\pm$0.63 & 21.2$\pm$0.66 & 21.4$\pm$0.69 & 21.9$\pm$0.64 & 7 \\
HOY J053518.12-050035.6 & 37.3$\pm$1.54 & 38.3$\pm$1.22 & 39.2$\pm$1.45 & 39.6$\pm$1.39 & 39.4$\pm$1.60 & 38.3$\pm$1.41 & 11 \\
HOY J053522.27-050116.8 & 67.4$\pm$1.34 & 67.9$\pm$1.29 & 71.5$\pm$1.47 & 70.3$\pm$1.42 & 72.7$\pm$1.63 & 72.8$\pm$1.50 & 12 \\
HHOY J053523.26-050132.6 & 151$\pm$1.97 & 156$\pm$2.07 & 159$\pm$2.20 & 160$\pm$2.25 & 158$\pm$2.58 & 166$\pm$2.13 & 4 \\
HOY J053525.89-050544.2 & 38.6$\pm$0.97 & 39.5$\pm$0.64 & 42.2$\pm$0.95 & 42.4$\pm$1.01 & 43.8$\pm$1.02 & 42.9$\pm$1.0 & 13 \\
HOY J053527.56-050934.5 & 269$\pm$9.30 & 279$\pm$9.35 & 281$\pm$9.42 & 282$\pm$9.28 & 289$\pm$8.63 & 274$\pm$8.69 & 11 \\
\enddata
\tablenotetext{a}{Peak-to-peak flux variations. \emph{Italic} values indicate sources that pass our variability criterion (see Section~\ref{sec:result} for details).}
\end{deluxetable}

\section{Results and Discussion}
\label{sec:result}

\begin{figure*}
  \centering
    \begin{tabular}{cc}
      \includegraphics[width=0.48\textwidth]{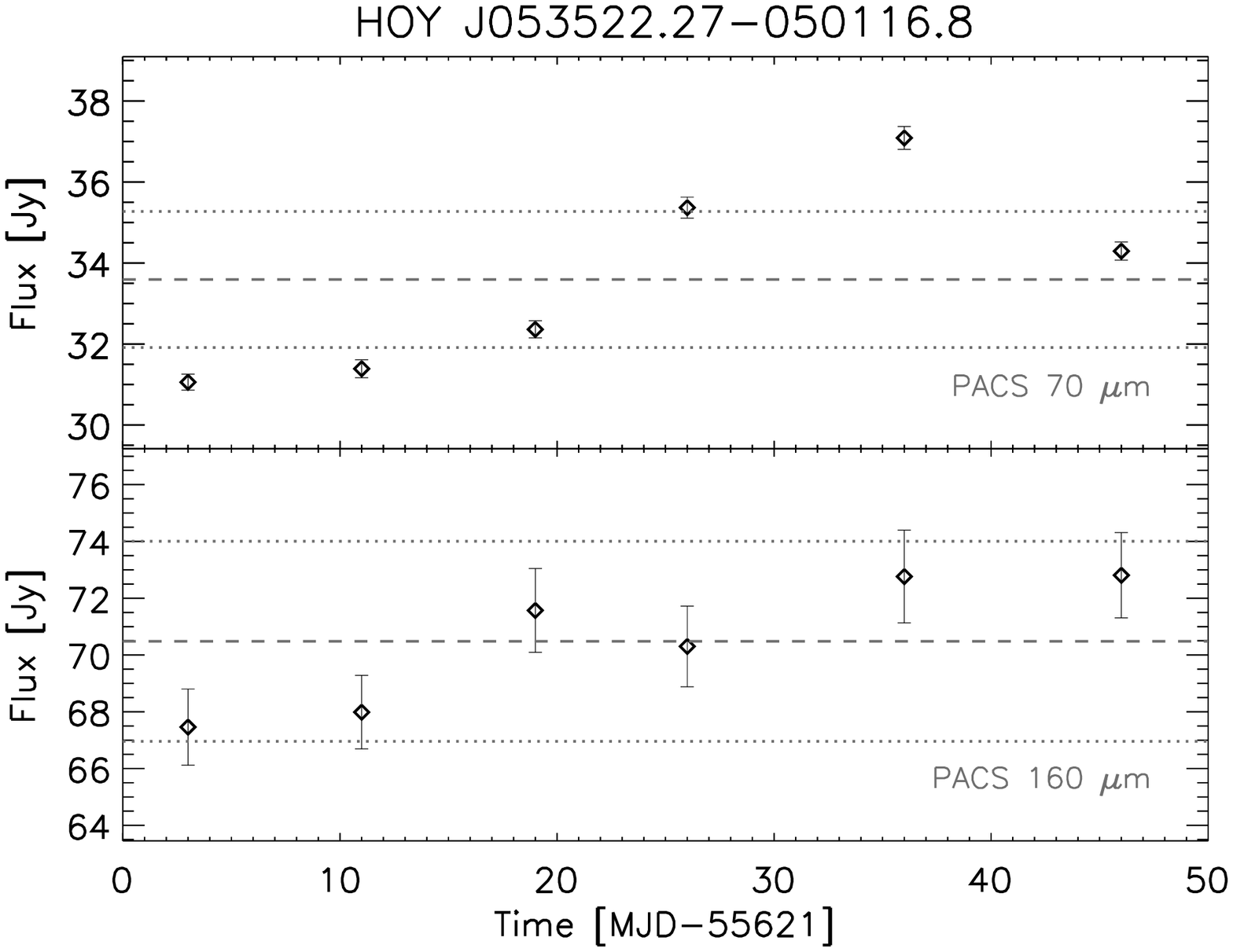}  & 
      	\includegraphics[width=0.48\textwidth]{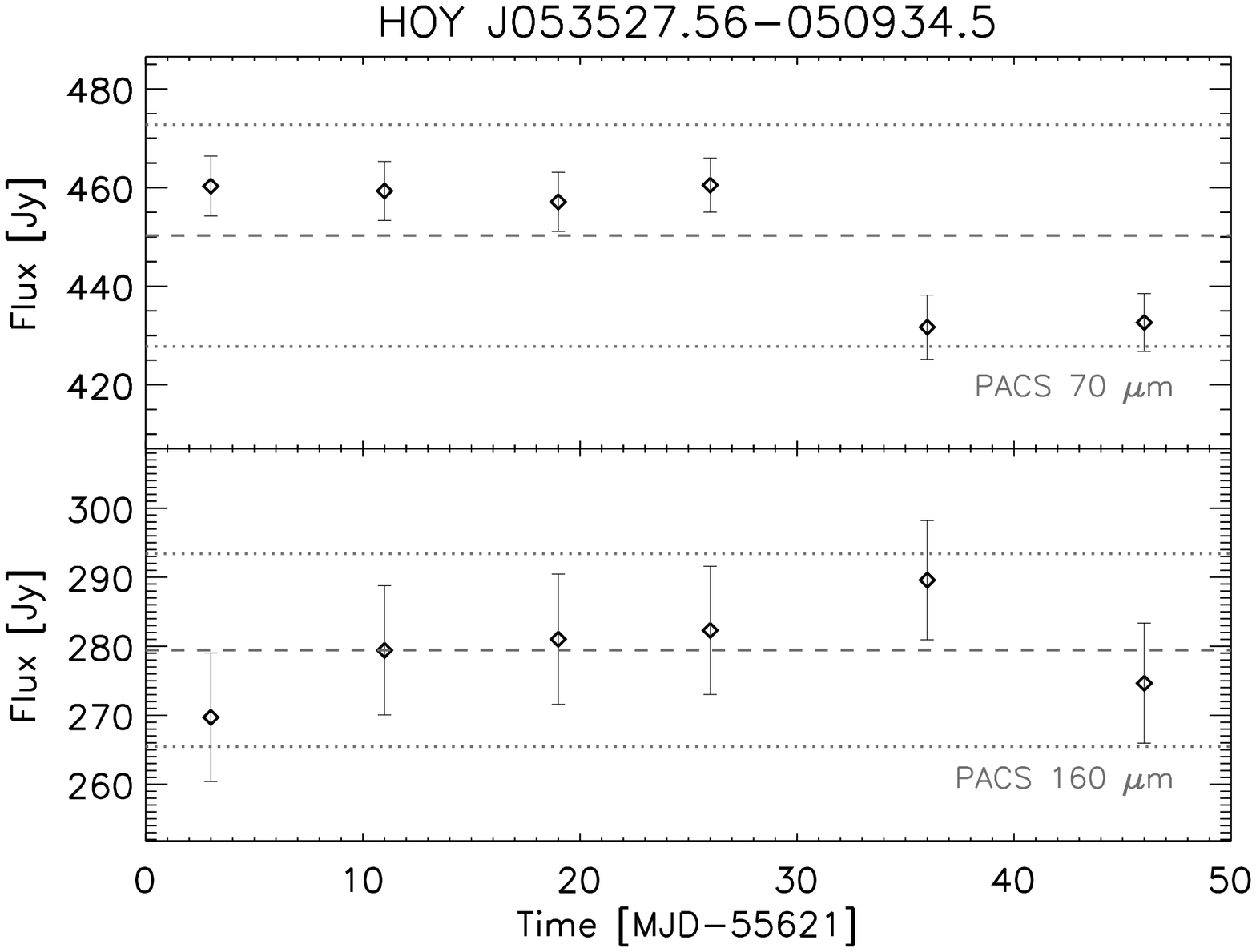} \\
      \includegraphics[width=0.48\textwidth]{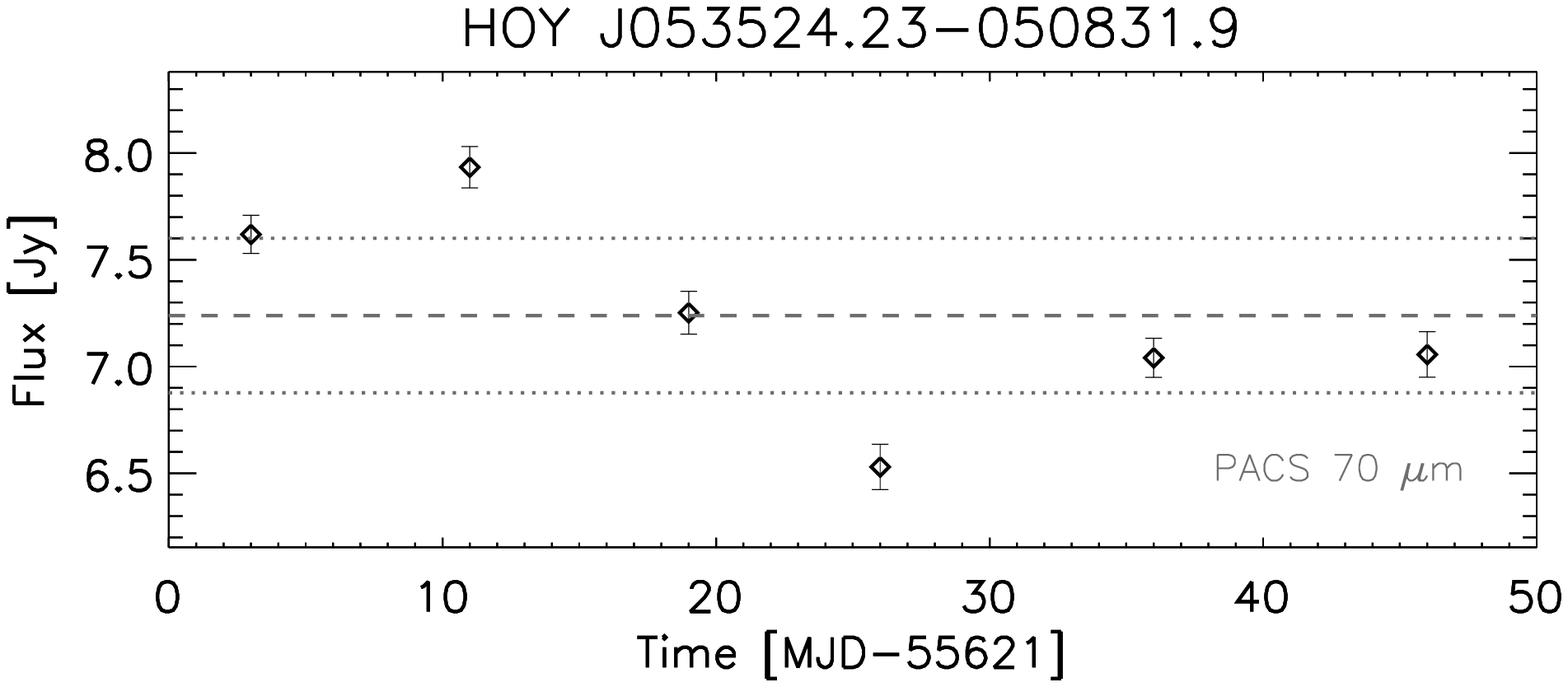}  & 
      	\includegraphics[width=0.48\textwidth]{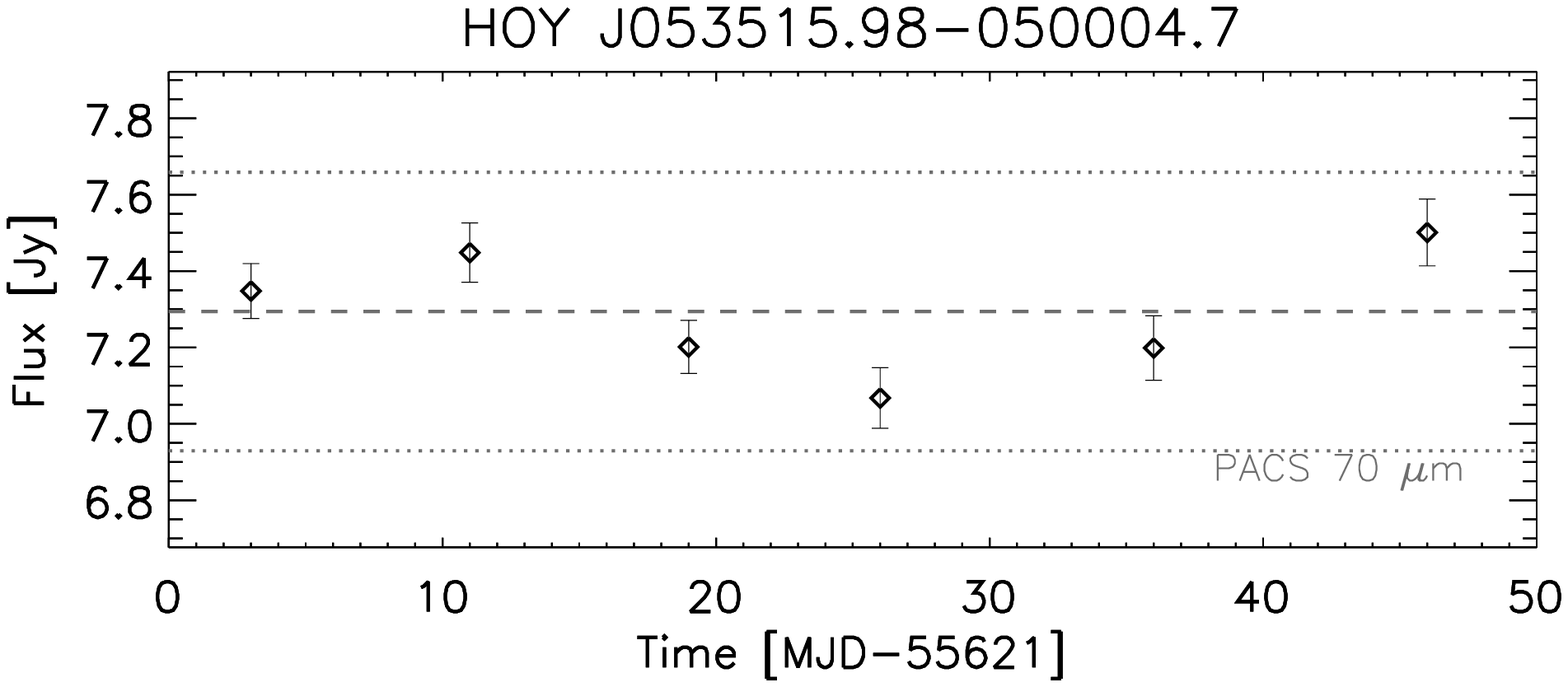} \\	
      	\includegraphics[width=0.48\textwidth]{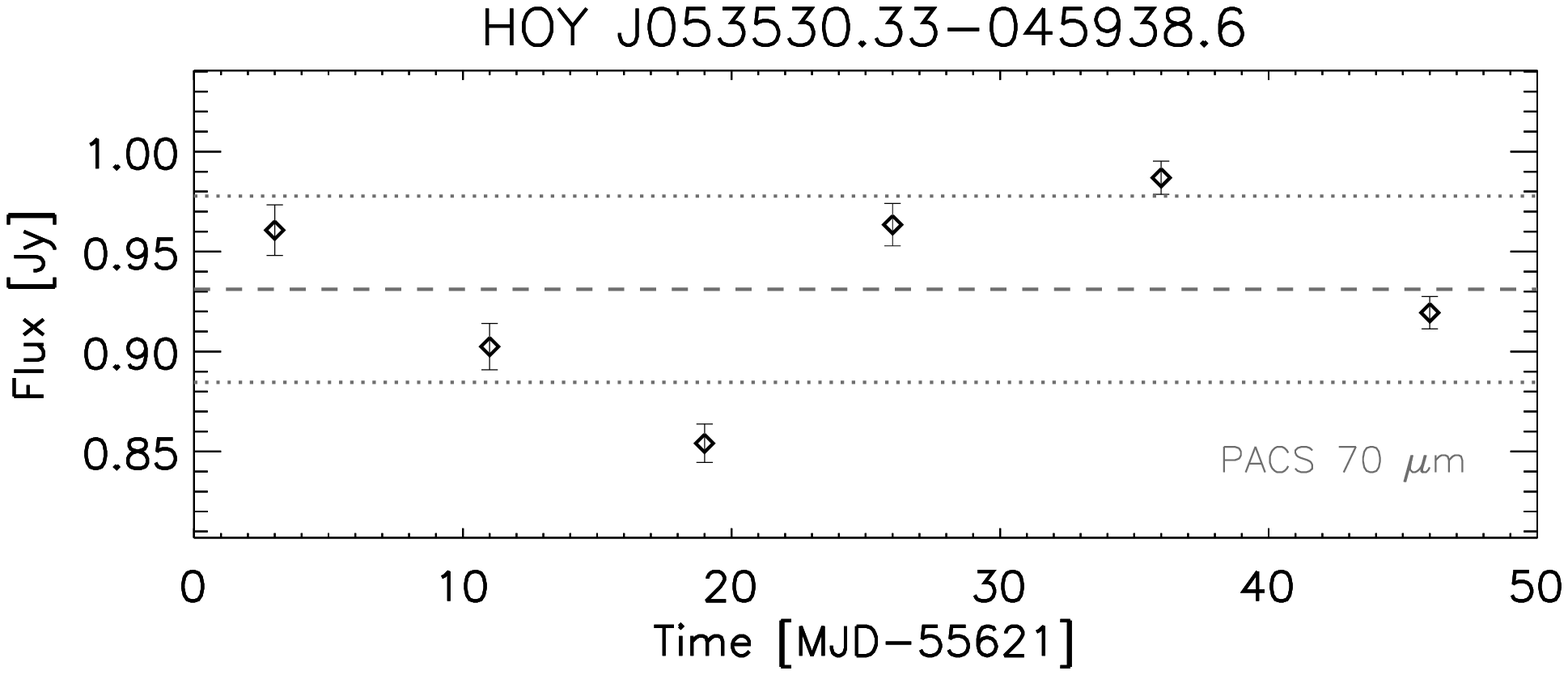} &	
      	\includegraphics[width=0.48\textwidth]{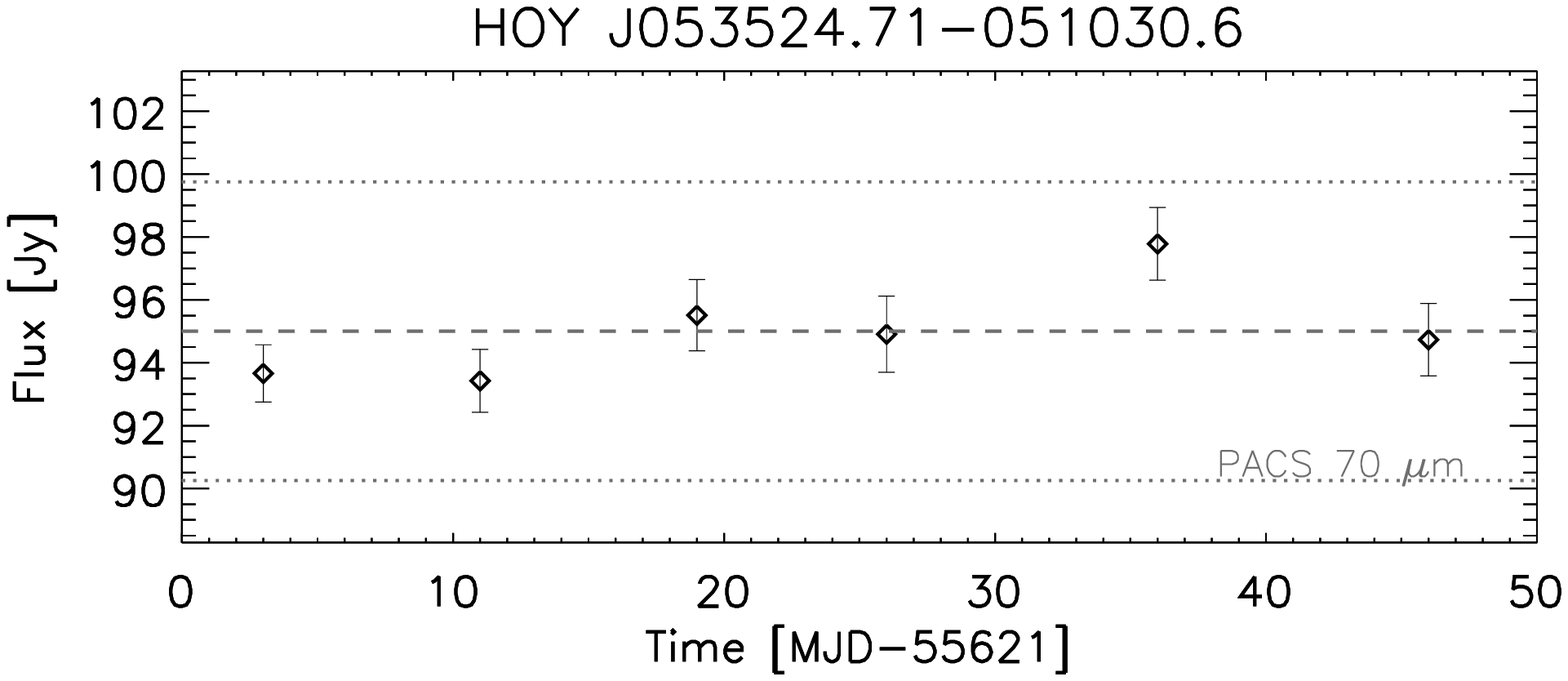} \\	
    \end{tabular}
  \caption{Sample of reliable PACS light curves drawn from Table~\ref{tab:fluxes_LC}. The left column presents the light curves of variable protostars, and the right column contains those that show flux variations within the estimated photometric uncertainties (see section~\ref{sec:photometry} for details). The set of graphs in the top row show the light curves at 70\,$\mu$m and 160\,$\mu$m when both are deemed reliable, while the other plots give 70\,$\mu$m fluxes only. The horizontal dashed and dotted lines give the average flux of the sources and the $\pm$5\% variations around the mean, respectively, indicating our level of confidence for variability detections. The first epoch was obtained on Feb. 26, 2011, corresponding to the Herschel Operational Day 653, or MJD~55618.}
    \label{fig:lightCurves}
\end{figure*}
\begin{figure*}
  \centering
    \begin{tabular}{cc}
      \includegraphics[width=0.48\textwidth]{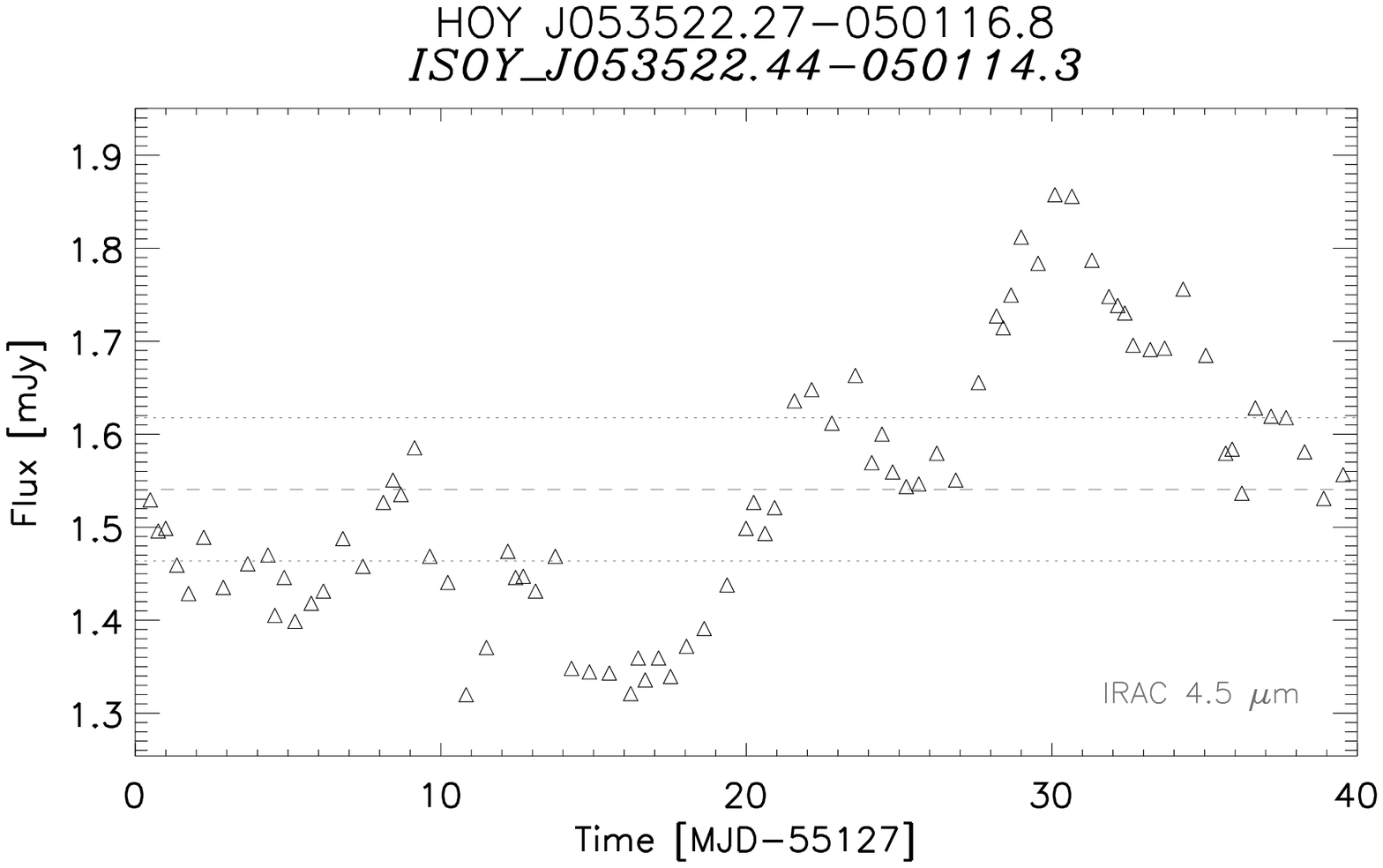}  & 
      	\includegraphics[width=0.48\textwidth]{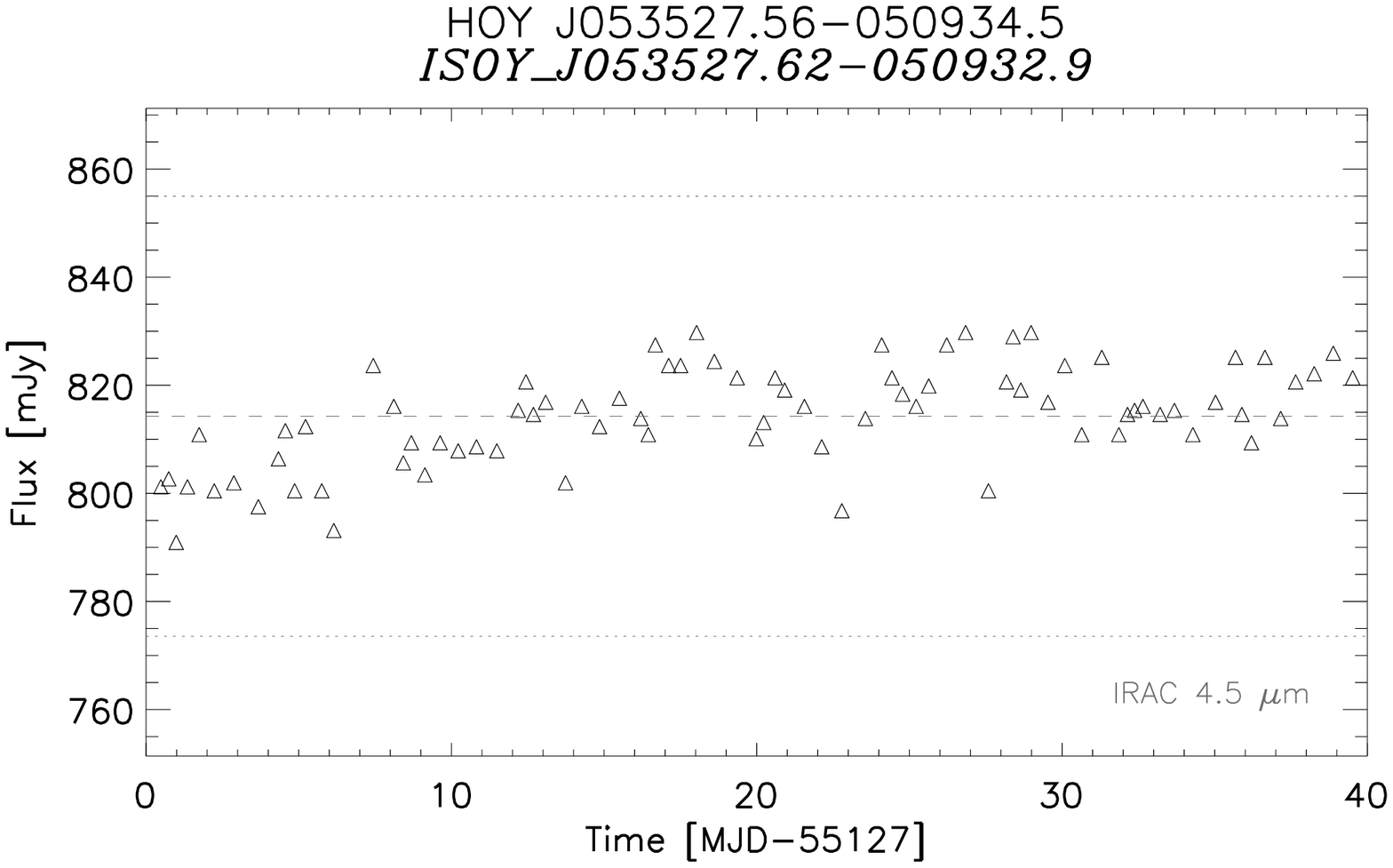} \\
      \includegraphics[width=0.48\textwidth]{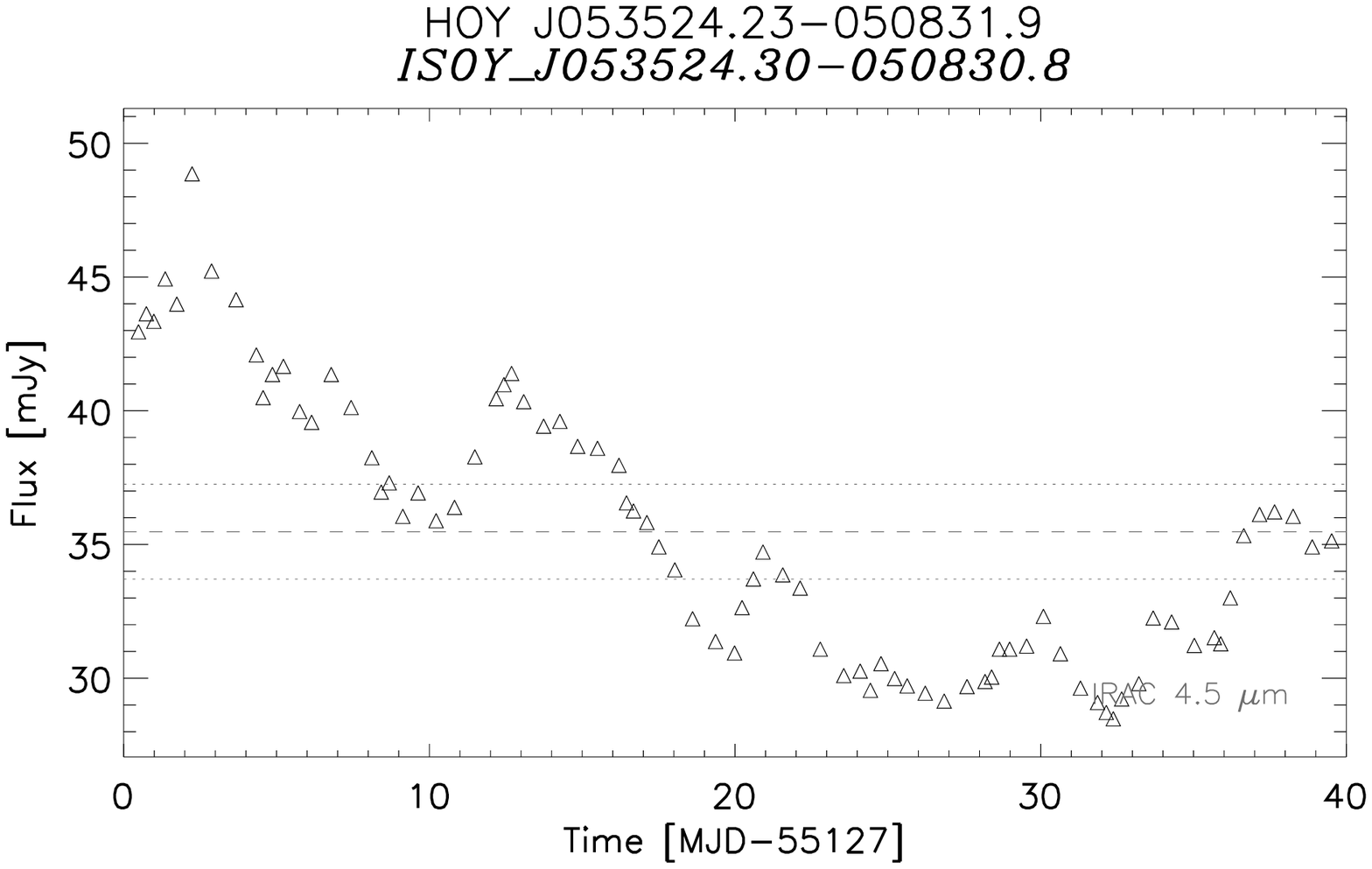}  & 
      	\includegraphics[width=0.48\textwidth]{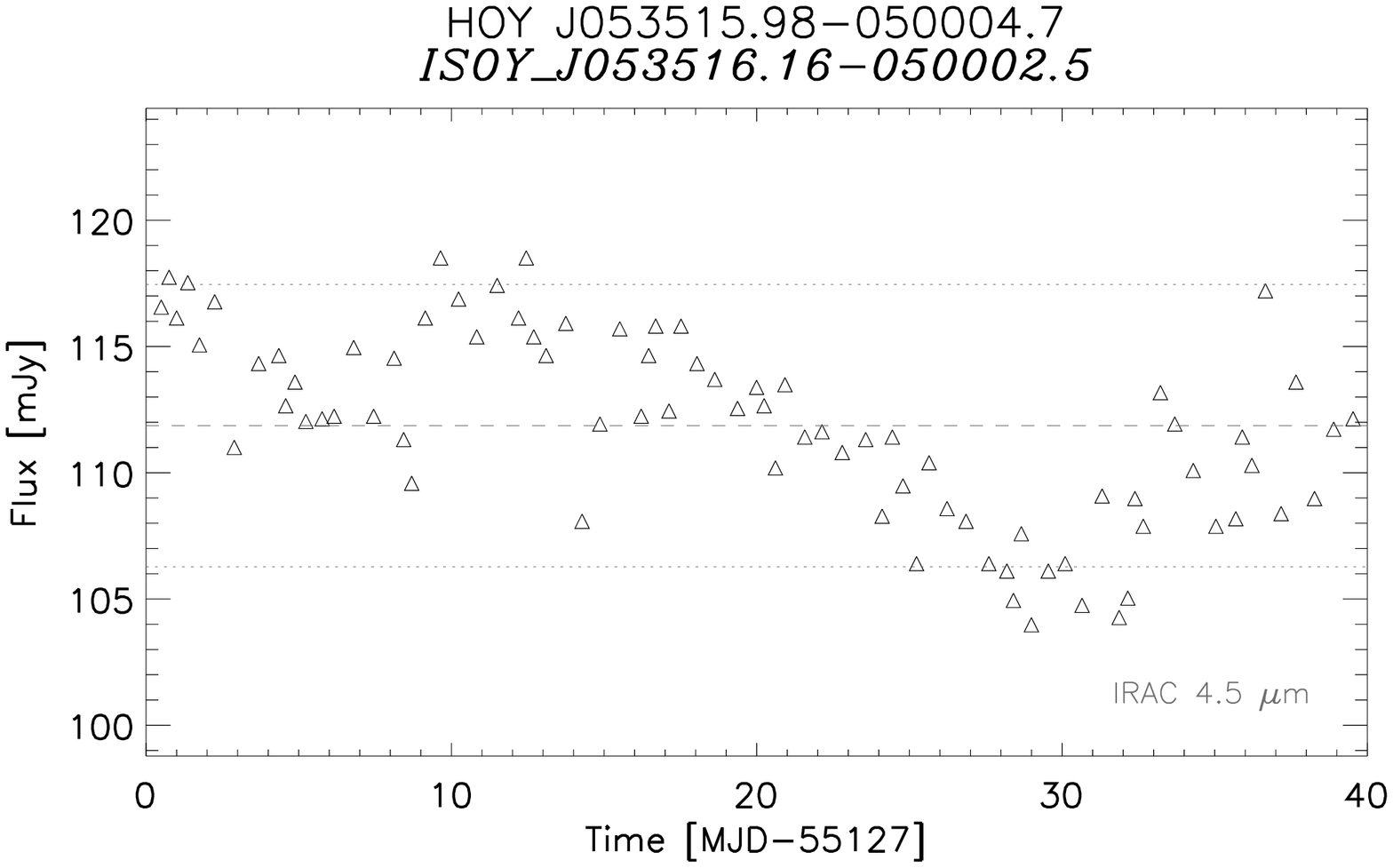} \\	
      	\includegraphics[width=0.48\textwidth]{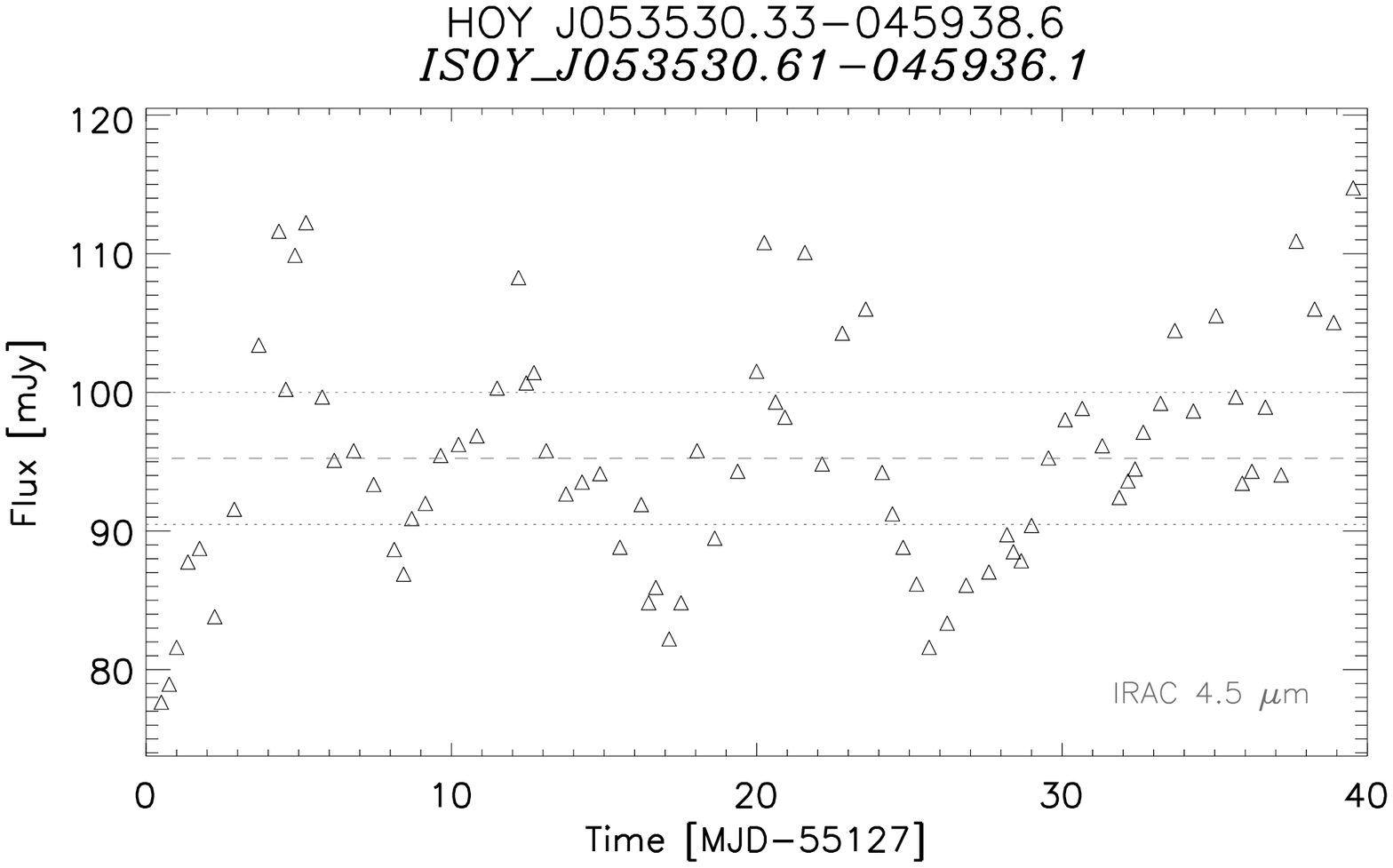} &	
      	\includegraphics[width=0.48\textwidth]{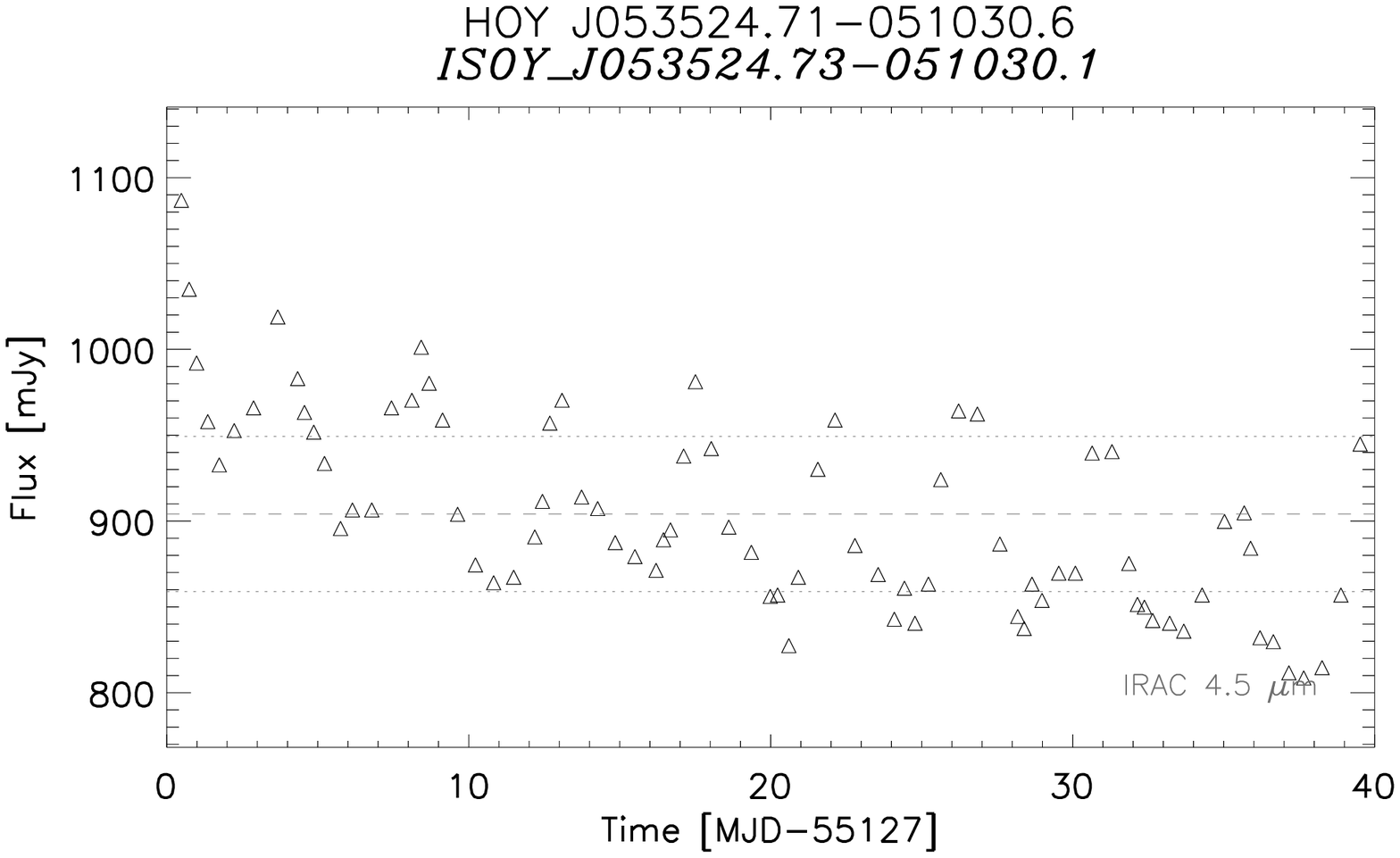} \\	
    \end{tabular}
  \caption{Spitzer/IRAC light curves measured at 4.5\,$\mu$m on the 6~objects presented in figure~\ref{fig:lightCurves}. These observations were obtained as part of the YSOVAR program in 2009, they are not contemporaneous with the Herschel data. The plots are arranged as in figure~\ref{fig:lightCurves} with the YSOVAR name given in the plot title, and the horizontal dashed and dotted lines also represent the mean and $\pm$5\% variations, respectively. }
    \label{fig:YSOVARlightCurves}
\end{figure*}

It is a difficult task to disentangle intrinsic source variability and photometric errors based solely on 6~data points, especially for faint sources.
Nevertheless, considering the criterions we used in selecting reliable light curves (cf section~\ref{sec:photometry}), we expect that flux variations greater than $\sim$10\% can be confidently attributed to the intrinsic variability of the observed sources.

We find that 8~sources out of 17~show peak-to-peak flux variations higher than $\sim$10\%. The left column of figure~\ref{fig:lightCurves} presents a sample of such light curves. The case of HOY$\,$J053522.27-050116.8 (top left plot) is of particular interest with its sine-like light curve that shows an amplitude of nearly 20\%. The smooth shape of the light curve suggests that the photometric uncertainties are very small for this source. In addition both 70$\,\mu$m and 160$\,\mu$m light curves show the same trend, which strengthens further our confidence that this source is indeed variable in the Far-IR. The object HOY$\,$J053524.23-050831.9 also shows strong variations in excess of 20\% in just over 2~weeks. For comparison, we present in figure~\ref{fig:YSOVARlightCurves} Spitzer light curves obtained in 2009 as part of the YSOVAR program \citep{morales11}, and it appears that objects variable in the far-IR tend to be also variable in the mid-IR (left column of figures~\ref{fig:lightCurves} and~\ref{fig:YSOVARlightCurves}). However the shape of Herschel and Spitzer light curves are not directly comparable since they are not contemporaneous\footnote{There was no significant overlap in the Spitzer and Herschel visibility windows of Orion in 2011.} and they have significantly different observing cadences.

The remaining 9~sources that have reliable light curves show flux variations in the 5 -- 10\% range. Although those sources exhibit variations below the 10\% threshold, we cannot exclude that they might as well be intrinsically variable in the far-IR. In fact, based on the Herschel dataset currently available, we are only sensitive to variations with timescales in the range 10 -- 50~days, and we miss all variations on shorter and longer timescales. For instance, the light curve of HOY~J053524.71-051030.6 does not satisfy our variability criterion according to its 70$\,\mu$m flux only, however its Spitzer light curve shows clear variability in the mid-IR with a period of $\sim$4~days (bottom right plot of figure~\ref{fig:YSOVARlightCurves}). If the mid- and far-IR variability had the same origin, and thus showed similar timescales at both wavelengths, then the sampling frequency of 10~days would not be appropriate to detect such rapid variations in Herschel bands. \\

Although YSOs have been monitored on daily to yearly timescales from the optical to the mid-IR and in all evolutionary stages \citep[e.g. ][]{herbst, morales11}, the variability parameter-space is still sparsely sampled in the far-IR. This is particularly true for high cadence monitoring and embedded protostars, as most far-IR monitoring programs have targeted outbursting YSOs with fading timescales of several years \citep[e.g. ][]{kospal}. The present Herschel time-series observations thus fill in this parameter-space gap, and they show that the far-IR emission, which is a good tracer of the internal luminosity of protostars \citep{dunham}, can vary noticeably on timescales as short as a couple of weeks. This is orders of magnitude shorter than the dynamical timescales of the far-IR emitting material that orbits at hundreds to thousands of AUs from the central protostar. Such short timescales indicate that the mechanism(s) responsible for the observed variability occurs on smaller spatial scales, either at the surface or within the disk, or close to the central protostar (r $\ll1\,$AU) where timescales are consistent with weekly to monthly timescales. 

Non-steady accretion likely plays a role in the observed variability. \citet[][and references therein]{zhu} argue that the typical accretion luminosity observed in protostars implies an accretion rate considerably lower than the predicted time-average infall rate in YSOs. This `luminosity problem' can be explained if infalling material accretes sporadically causing major accretion events that are sufficiently short-lived so that protostars are usually observed in quiescence. Numerical models of \citet{vorobyov} manage to reproduce these accretion events for Class\,I and Class\,II YSOs when disk fragments fall onto the central protostar. Such energetic outbursts have been observed in FU-Orionis-type objects with typical timescales of 1 -- 10~years \citep{hartmann}. Similar events might occur on shorter temporal and less energetic scales. 

For instance, \citet{flaherty11} measured accretion rates from Pa$\beta$ and Br$\gamma$ observations, and they find that the accretion rate of the Class$\,$II object~LRLL~31 varies by a factor of five, with the largest changes occurring on a weekly timescale. These episodic accretion events could be pictured as knotty filaments of gas funneling from the inner region of the circumstellar disk to the central star, likely through magnetic field lines \citep{shu}, and releasing packets of energy as the clumps hit the surface of the protostar. For embedded protostars, the energetic photons generated in the accretion shock are reprocessed by the dense and dusty surrounding envelop, and they eventually escape the protostar in the far-IR regime. The accretion luminosity variations that originate from episodic accretion events in the central region of a protostar propagate outwards and temporarily warm up the envelop, which in turn leads to detectable far-IR flux variations. The typical timescale for photons to reach the far-IR emitting material in the envelop (located at $\sim10^2-10^3\,$AU from the central star) is roughly the light travel time, which is a few days, similar to the typical timescales observed in the light curves of Figure~\ref{fig:lightCurves}.


The alternative scenario proposed by \citet{flaherty} might also cause a measurable far-IR variability in YSOs. It consists in a variable scale height of the disk inner edge that casts a shadow on the outer disk, thus cooling down the mid- and far-IR-emitting disk. This non-axisymetric disk model was used to reproduce the mid-IR variability observed in the YSO LRLL~31, and it can potentially give constraints on the inner disk structure. In a subsequent work, \citet{flaherty11} have tested various models that may lead to variable scale heights of the inner edge of the disk, with origins ranging from variable accretion, perturbations by a companion, winds, and the influence of magnetic fields. Although LRLL~31 is more evolved than most objects observed in the present study, a similar mechanism could well explain the variations of the Class\,II source HOY\,J053530.33--045938.6 (bottom left plot in figure~\ref{fig:lightCurves}). However, contemporaneous near-IR observations would be necessary to confirm the expected anti-correlated flux variations with respect to the far-IR light curves.\\

The first 6~epochs of our monitoring program have demonstrated that protostellar emission can vary on relatively short timescales in the far-IR. Furthermore we find that the fraction of variable protostars in our sample is relatively high ($>$40\%), though difficult to ascertain based on sparse 70$\,\mu$m light curves only. For a better sampling of the light curves, we have requested a higher observing cadence for the second Orion visibility window, and we have extended the initial span of the monitoring program with the last observations scheduled for autumn~2012, thus covering a period of nearly 2~years. These additional observations, with refined photometric measurements and a detailed SED modeling, should increase the number of variability detections and help us understand the nature of these objects and the mechanism responsible for the observed variability. A spectroscopic follow-up with Herschel/PACS is also scheduled for 2012 to monitor accretion activity in a couple of variable sources with atomic and molecular lines, particularly [OI], [CII], CO and H$_2$O, already observed as part of the HOPS program \citep{megeath,manoj}.\\


This work is based on observations made with Herschel, a European Space Agency Cornerstone Mission with significant participation by NASA. Support for this work was provided by NASA through an award issued by JPL/Caltech. This research has made use of NASA's Astrophysics Data System.

{\it Facilities:} \facility{Herschel Space Observatory (PACS)}.

\clearpage


\begin{thebibliography}{}

\bibitem[Billot et al.(2006)]{billot06} Billot, N. et al., 2006, SPIE Proc., 6265
\bibitem[Billot et al.(2010)]{billot10} Billot, N. et al., 2010, SPIE Proc., 7741
\bibitem[Cantalupo et al.(2010)]{cantalupo} {Cantalupo}, C.~M., {Borrill}, J.~D., {Jaffe}, A.~H., {Kisner}, T.~S. and {Stompor}, R., 2010, \apjs, 187, 212
\bibitem[Carpenter et al.(2001)]{carpenter} {Carpenter}, J.~M., {Hillenbrand}, L.~A. and {Skrutskie}, M.~F., 2001, \aj, 121, 3160
\bibitem[Dunham et al.(2008)]{dunham} {Dunham}, M.~M. et al., 2008, \apjs, 179, 249
\bibitem[Espaillat et al.(2011)]{espaillat} {Espaillat}, C., {Furlan}, E., {D'Alessio}, P., {Sargent}, B., {Nagel}, E., {Calvet}, N., {Watson}, D.~M. and {Muzerolle}, J., 2011, \apj, 728, 49
\bibitem[Flaherty \& Muzerolle(2010)]{flaherty} Flaherty, K.~M. \& {Muzerolle}, J., 2010, \apj, 719, 1733
\bibitem[Flaherty et al.(2011)]{flaherty11} Flaherty, K.~M. et al., 2011, \apj, 732, 83
\bibitem[Greene et al.(1994)]{greene} {Greene}, T.~P., {Wilking}, B.~A., {Andre}, P., {Young}, E.~T. and {Lada}, C.~J., 1994, \apj, 434, 614
\bibitem[Hartmann \& Kenyon(1996)]{hartmann} Hartmann, L. \& {Kenyon}, S.~J., 1996, \araa, 34, 207
\bibitem[Harvey et al.(1998)]{harvey} {Harvey}, P.~M., {Smith}, B.~J., {di Francesco}, J. and {Colome}, C., 1998, \apj, 499, 294
\bibitem[Herbst et al.(2000)]{herbst} {Herbst}, W., {Rhode}, K.~L., {Hillenbrand}, L.~A. and {Curran}, G., 2000, \aj, 119, 261
\bibitem[Joy(1945)]{joy} Joy, A.~H., 1945, \apj, 102, 168
\bibitem[Juh{\'a}sz et al.(2007)]{juhasz} {Juh{\'a}sz}, A., {Prusti}, T., {{\'A}brah{\'a}m}, P. and {Dullemond}, C.~P., 2007, \mnras, 374, 1242
\bibitem[Ke et al.(2012)]{ke} Ke, T.~T., {Huang}, H., and {Lin}, D.~N.~C., 2012, \apj, 745, 60
\bibitem[K{\'o}sp{\'a}l et al.(2007)]{kospal} {K{\'o}sp{\'a}l}, {\'A}., {{\'A}brah{\'a}m}, P., {Prusti}, T., {Acosta-Pulido}, J., {Hony}, S., {Mo{\'o}r}, A. and {Siebenmorgen}, R., 2007, \aap, 470, 211
\bibitem[Kryukova et al.(2012)]{kryukova} {Kryukova}, E. and {Megeath}, S.~T. and {Gutermuth}, R.~A. and {Pipher}, J. and {Allen}, T.~S. and {Allen}, L.~E. and {Myers}, P.~C. and {Muzerolle}, J., 2012, arXiv, 1204.1535
\bibitem[Lemke et al.(1996)]{lemke} Lemke, D. et al., 1996, \aap, 315, 64
\bibitem[Megeath et al.(2011)]{megeath} Megeath, T. et al., 2011, IAU Symposium, 280, 254
\bibitem[Megeath et al.(2012)]{megeath12} Megeath, T. et al., 2012, Submitted to ApJ
\bibitem[Morales-Calder{\'o}n et al.(2009)]{morales09} {Morales-Calder{\'o}n}, M. et al., 2009, \apj, 702, 1507
\bibitem[Morales-Calder{\'o}n et al.(2011)]{morales11} {Morales-Calder{\'o}n}, M. et al., 2011, \apj, 733, 50
\bibitem[Muzerolle et al.(2009)]{muzerolle} {Muzerolle}, J. et al., 2009, \apjl, 704, 15
\bibitem[Ott (2010)]{ott} {Ott}, S. 2010, ASP Conference Series, 434, 139
\bibitem[Pilbratt et al.(2010)]{pilbratt} Pilbratt, G.~L. et al., 2010, \aap, 518, 1
\bibitem[Poglitsch et al.(2010)]{poglitsch} Poglitsch, A. et al., 2010, \aap, 518, 2
\bibitem[Puravankara et al.(2011)]{manoj} Manoj Puravankara et al., 2011, American Astronomical Society Meeting Abstracts \#217, 43, 255.04
\bibitem[Roussel et al.(2012)]{roussel} Roussel, H. et al., 2012, arXiv:1205.2576v1
\bibitem[Shu et al.(1994)]{shu} {Shu}, F. et al., 1994, \apj, 429, 781
\bibitem[Sitko et al.(2008)]{sitko} Sitko, M.~L. et al., 2008, \apj, 678, 1070
\bibitem[Turner et al.(2010)]{turner} {Turner}, N.~J., {Carballido}, A. and {Sano}, T., 2010, \apj, 708, 188
\bibitem[Vorobyov \& Basu(2010)]{vorobyov} {Vorobyov}, E.~I., and {Basu}, S., 2010, \apj, 719, 1896
\bibitem[Vrba et al.(1986)]{vrba} {Vrba}, F.~J., {Rydgren}, A.~E., {Chugainov}, P.~F., {Shakovskaia}, N.~I. and {Zak}, D.~S., 1986, \apj, 306, 199
\bibitem[Werner et al.(2004)]{werner} Werner, M.~W. et al., 2004, \apjs, 154, 1
\bibitem[Zhu et al.(2009)]{zhu} Zhu, Z., Hartmann, L. \& Gammie, C., 2009, \apj, 694, 1045
  
\end{thebibliography}
\end{document}